\documentclass[nofootinbib,twocolumn,aps,pre,superscriptaddress,citeautoscript,floatfix]{revtex4-2}
\usepackage{graphics}
\usepackage{graphicx}
\usepackage{color}
\usepackage{amsmath}
\usepackage{amssymb}
\usepackage{float}

\newcommand{\be}{\begin{equation}} 
\newcommand{\ee}{\end{equation}}
\newcommand{\bea}{\begin{eqnarray}}
\newcommand{\eea}{\end{eqnarray}}
\newcommand{\D}{{\rm d}}
\newcommand{\G}{{\cal K}} 
\newcommand{\kB}{k_{\rm B}}
\newcommand{\E}{{\rm e}}
\def\bnabla{\mbox{\boldmath $\nabla $}}

\begin{document}
\author{Du Chen}
\affiliation{Department of Physics, Wenzhou University, Wenzhou, Zhejiang 325035, China}
\affiliation{Zhejiang Key Laboratory of Soft Matter Biomedical Materials, Wenzhou Institute, 
University of Chinese Academy of Sciences, Wenzhou, Zhejiang 325001, China}
\author{Rudolph Podgornik}
\thanks{Deseased December 28, 2024}
\affiliation{School of Physical Sciences, University of Chinese Academy of Sciences, Beijing 100049, China}
\affiliation{Zhejiang Key Laboratory of Soft Matter Biomedical Materials, Wenzhou Institute, 
University of Chinese Academy of Sciences, Wenzhou, Zhejiang 325001, China}
\author{David Andelman}
\affiliation{School of Physics and Astronomy, \& Center for Physics and Chemistry of Living Systems,\\
Tel Aviv University, Ramat Aviv 69978 Tel Aviv, Israel}
\author{Xianghong Wang}\thanks{wangxianghong@stiei.edu.cn}
\affiliation{School of Sino-German Engineering, Shanghai Technical Institute of Electronics and Information,\\ Shanghai 201411, China}
\author{Linli He}\thanks{linlihe@wzu.edu.cn}
\affiliation{Department of Physics, Wenzhou University, Wenzhou, Zhejiang 325035, China}
\author{Shigeyuki Komura}\thanks{komura@wiucas.ac.cn}
\affiliation{Zhejiang Key Laboratory of Soft Matter Biomedical Materials, Wenzhou Institute, 
University of Chinese Academy of Sciences, Wenzhou, Zhejiang 325001, China}
\author{Bin Zheng}\thanks{email: zhengbin@wiucas.ac.cn}
\affiliation{Zhejiang Key Laboratory of Soft Matter Biomedical Materials, Wenzhou Institute, 
University of Chinese Academy of Sciences, Wenzhou, Zhejiang 325001, China}

\title{Non-uniform swelling of polyelectrolyte hydrogels: effects of charge regulation}

\begin{abstract}
We investigate the impact of charge regulation (CR) on the non-uniform swelling behavior of polyelectrolyte hydrogels.
The Poisson-Boltzmann theory with electro-elastic coupling between the local polymer density and elastic deformation is considered.
We investigate the spatial distributions of the elastic displacement and polymer density 
under different salt concentrations and compare charge-regulated gels with fixed-charge (non-CR) gels of the same net charge. 
Our results show that the CR induces spatially varying charge fractions, 
which strengthen the electro-elastic response and lead to stronger non-uniform swelling compared with non-CR gels. 
These findings provide a theoretical basis for understanding and controlling non-uniform swelling in responsive polyelectrolyte hydrogels.
\end{abstract}
	
\maketitle
	
\section{Introduction}
%
Polyelectrolyte (PE) hydrogels are an essential class of soft matter materials and have been widely used in various applications, 
including drug delivery, water purification, and biomimetic actuators for soft robotics 
systems~\cite{Li2016,Guo2020,Jeon2017,Doring2013}. A key feature underlying 
many of them is their pronounced stimulus-responsive volume change. In weakly charged gels, the swelling reflects 
a competition between the electrostatic repulsion between charged monomers, osmotic pressure, and the gel network elasticity. 
Moreover,  many PE networks carry weakly dissociable acidic or basic moieties, whose ionization state varies with solution 
pH or salt concentration. Therefore, the gel volume can be effectively regulated by the surrounding environment.

Gel swelling was studied by T. Tanaka and his coworkers~\cite{Tanaka1984,Tokita2022}, who explained the continuous and discontinuous 
volume phase transitions in the framework of Donnan equilibrium theory. 
Subsequent works incorporated more explicit electrostatics through the linearized Debye-H\"uckel description of the full 
non-linear Poisson-Boltzmann formulation coupled with the Flory-Huggins interaction and network 
elasticity~\cite{English1998,Prausnitz2006,Levin2014,Nikam2020,Monica2011}, 
as well as extensive particle-based simulations~\cite{Beyer2022,Jonas2019}. 

Later on, {\it charge regulation} (CR) effect, {\it i.e.}, the association-dissociation 
equilibria leading to adaptive ionization~\cite{ninham1971electrostatic} 
has been applied to gels and weak polyelectrolytes~\cite{Lund_2013, markovich2021charged, avni2019charge}. 
Muthukumar and coworkers~\cite{Muthu2010, Muthu2012} introduced a self-consistent {\it charge regularization} 
that accounts for the dynamic variation of the network charge in response to changes in solvent pH and 
salt concentration. This term is identical to what others call ``charge regulation" 
(CR). Zheng {\it et al.}~\cite{zheng2023} further demonstrated that CR can induce a discontinuous volume phase transition in gel swelling. 

Recently, gel swelling was shown not only to depend 
on the amount of charge but also on the charge spatial distribution 
and its effect on network elasticity~\cite{Muthu2012, zheng2023, Nermin2017,Sooraj2022}. More specifically, the elastic response 
can be explicitly tuned by the charge spatial distribution~\cite{Nermin2017,Sooraj2022}. CR adjusts 
ionization equilibria in response to changes in pH and salt concentration, thereby altering the charge distribution and modifying 
the elastic response. Understanding how CR affects non-uniform swelling in polyelectrolyte gels via this process remains 
limited and largely unresolved.

In the present work, we propose a theoretical framework to study the effects of CR on the {\it non-uniform} swelling of the PE gel. 
The coupling between the local polymer density and the local elastic displacement is considered.
Based on the Poisson-Boltzmann theory and the free energy of a two-site CR 
model~\cite{zheng2021,zheng2023,zheng2025}, we derive the coupled equations 
for the electrostatic potential and elastic displacement. We then analyze the electro-elastic response 
in a free-energy framework and find that the CR has a significant effect on the elastic displacement. It also enhances non-uniform swelling, 
as shown by direct comparison with the corresponding fixed-charge (non-CR) case. 

The outline of our paper is as follows. Section~II contains the model description with 
its free energy and boundary conditions. In Sec.~III, we present the results, 
addressing the variations in the electrostatic potential, elastic displacement, and the spatial variation 
of the gel density. 
We analyze the relationship between the gel size and salt concentration, and 
discuss the CR effect by comparing it with the non-CR case. 
Finally, in Sec.~IV, we present the discussion and conclusions.


\begin{figure}[!t]
	{\includegraphics[width=0.35\textwidth,draft=false]{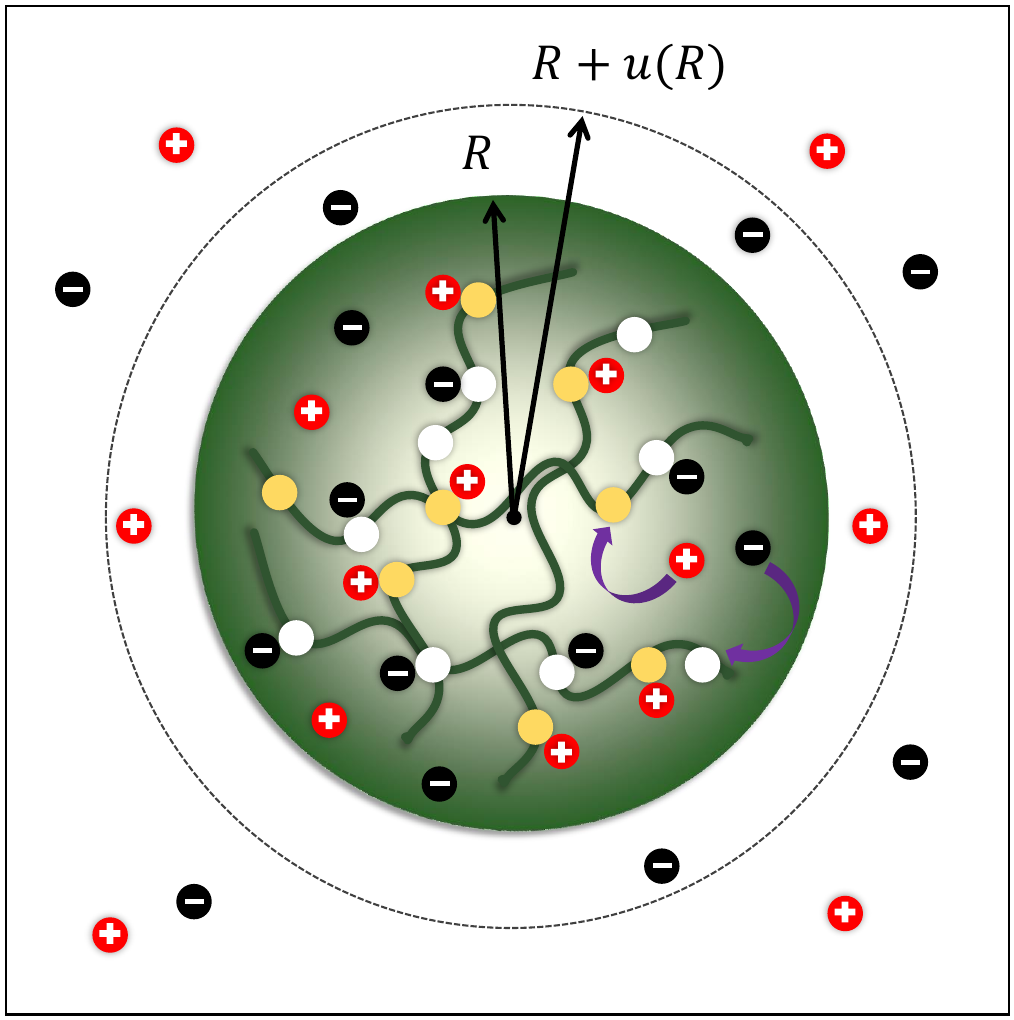}}
	\caption{
		Schematic drawing of the CR gel model. The green sphere represents the gel volume. 
		In the two-site model, the polymer chains contain n-type and p-type sites that can become 
		negatively (white sites can be associated with black anions) 
		or positively charged (yellow sites can be associated with red cations) through an association/dissociation process. 
		The original gel radius is $R$, while $R+u(R)$ is the swollen gel radius. Free cations are marked in red and anions in black.
	}
	\label{fig1}
\end{figure}

\section{Model and free energy}

Our system is composed of a spherical polyelectrolyte (PE) gel of radius $R$ placed in a 1:1 monovalent ionic solution 
containing ions with  bulk densities (per unit volume) 
$n_{\pm}^{\rm b}\,{=}\,n_{\rm b}$, as shown in Fig.~\ref{fig1}. 
The gel contains $N_{\rm c}$ polymer chains of local chain number density $c_{\rm p}({\bf r})$ (per unit volume), where 
${\bf r}$ denotes the position vector measured from the gel center. 
The average polymer chain number density over the gel volume in the undeformed state (radius $R$) is denoted by $\bar{c}_{\rm p}$.

Each polymer chain contains two types of ionizable sites, p-type and n-type.
The p-type sites (colored in yellow in Fig.~\ref{fig1}) can become positively charged upon cation adsorption.
The corresponding adsorption/desorption is described by the following chemical reaction
\be
{\rm p} + {\rm C}^{+}\rightleftharpoons {\rm pC}^{+},
\label{chemAC}
\ee
whereas the n-type sites (colored white in Fig.~\ref{fig1}) 
can become negatively charged through the adsorption of an anion, and similarly is described by the chemical reaction
\be
{\rm n}+{\rm D}^{-}\rightleftharpoons {\rm nD}^{-}.
\label{chemBD}
\ee

More specifically, each polymer chain consists of $N$ monomers; out of them $N_{+}\,{=}\,\gamma_{+} N$ 
are p-type and $N_{-}\,{=}\,\gamma_{-} N$ are n-type. The parameters $\gamma_\pm\,{=}\,N_\pm/N $ 
are defined as the overall fractions of chain sites that can undergo the CR process, while $0\leq \phi_\pm({\bf r})\leq 1$ 
denote the fractions of the corresponding ionizable sites that are charged at position $\bf r$. 
Accordingly, $\phi_\pm({\bf r})=0$ corresponds to the limit in which none of the 
$N_\pm$ sites are charged, while $\phi_\pm({\bf r})=1$ indicates that all the $N_\pm$ sites are charged.

We consider a radially symmetric gel. Therefore, $\phi_\pm$ as well as the gel density, 
vary only along the radial coordinate, $r\,{=}\,\lvert {\bf r} \rvert$.
The equilibrium values of these quantities are determined by minimizing the free energy, as explained below.

The total free energy of the system can be written as a sum of five terms,
\bea
F=\int {\rm d}V \,f &=& \int {\rm d}V \bigg[ f_{\rm elec} +f_{\rm mix} + f_{\rm elas} \nonumber\\
&+& c_{\rm p}({\bf r}) f_{\rm CR}  - \sum_{i = \pm}  \mu_i n_i({\bf r})  \bigg],
\label{f_tot} 
\eea
where $\mu_\pm$ are the chemical potentials of the mobile cations/anions, and $n_\pm({\bf r})$ are their 
number densities (per unit volume). 
The first term, $f_{\rm elec}$, is the electrostatic contribution, expressed as
\bea
f_{\rm elec} &=& - {\frac12} \varepsilon (\bnabla \psi)^2 \ +\  \psi \sum_{i = \pm} q_i n_i 
\nonumber\\
&-& e \psi c_{\rm p}({\bf r}) {\mathcal H}(R-r)( N_{-} \phi_{-} -N_{+} \phi_{+}),
\eea
where $\varepsilon\,{=}\,\varepsilon_0\varepsilon_r$ is the dielectric constant of the aqueous solution, 
$\varepsilon_0$ is the vacuum permittivity, $\varepsilon_r \approx 80$ is the water relative dielectric constant, 
$\psi$ is the electrostatic potential, and the cation/anion mobile species have charge $q_\pm$. 
In this study, for simplicity's sake, we consider only monovalent salts with $q_\pm\,{=}\,\pm e$ and 
$e$ being the elementary charge, and $n_\pm^{\rm b}=n_{\rm b}$. Finally, the  
Heaviside function used above is defined as usual as 
\bea
{\mathcal H}(x)=
\begin{cases}
	0  & x<0,\\
	1  & x\geq 0.
\end{cases}
\eea 

The second term in Eq.~(\ref{f_tot}), $ f_{\rm mix}$ is the mixing free energy of the $\pm$ charged species  
assumed to have the form of an ideal gas entropic contribution,
\be
f_{\rm mix} =  \kB T \sum_{i=\pm} \Big[ n_i \ln(n_i a^3) - n_i \Big],
\label{nuiop1}
\ee
where $\kB$ is the Boltzmann constant and $T$ is the temperature. 
All ionic species are assumed to have the same molecular volume $a^3$.
The exchange between mobile ions and the gel association/dissociation 
sites is described via the Langmuir isotherm, which is consistent with the chemical ionic reactions, 
Eqs.~\eqref{chemAC}-\eqref{chemBD}. Hence,
the CR free energy per polymer chain in Eq.~(\ref{f_tot}) is given by
\bea
f_{\rm CR} &= &{\mathcal H}(R-r) \kB T \Big[-\sum_{i = \pm} N_i\phi_i \,A_i  
\nonumber \\
&+&  \sum_{i = \pm}{N_i} \Big[ \phi_i \ln{\phi_i} 
+  (1-\phi_i )\ln{(1-\phi_i)}   \Big]\Big],
\label{VO22}
\eea
where 
\be
A_\pm\equiv \frac{\alpha_\pm  + \mu_\pm}{\kB T}
\label{Apm}
\ee
and $\alpha_\pm$ are the association/dissociation energy parameters for a single p- or n-type site.

The elastic term in Eq.~(\ref{f_tot}), $f_{\rm elas}$, is written by assuming that the gel is an isotropic linear elastic medium. 
Hence, the elastic free-energy density is expressed as
\bea
f_{\rm elas}({\bf u}) &=& \frac{\G}{2} (\bnabla \cdot {\bf u})^2 
+ \frac{\bar{\mu}}{2}  (\bnabla \times {\bf u})^2,
\eea
where $\G \,{=}\, \bar{\lambda} +2\bar{\mu}$,  $\bar{\lambda}$ and $\bar{\mu}$
are the Lam\'e elastic coefficients, and $ {\bf u}$ is the elastic displacement vector.
Since we deal only with fields that have spherical symmetry, the second term vanishes: $\bnabla \times {\bf u} \,{=}\, 0$.

We include a coupling between the gel's local deformation and polymer chain density. 
Under deformation, local conservation of the polymer chains requires
\be
c_{\rm p}({\bf r})=\frac{\bar{c}_{\rm p}}{\det({\bf I}+\bnabla{\bf u})}, 
\ee
where $\bnabla{\bf u}$ 
is the displacement gradient tensor and ${\bf I}$ is the identity tensor. 
The local volume change ratio associated with the deformation~\cite{Mannattil2025} 
is $\det({\bf I}+\bnabla{\bf u})$, where $\det(...)$ denotes the determinant of a second-rank tensor.
For small deformations,
\be
\det({\bf I}+\bnabla{\bf u})\simeq1+\bnabla\cdot{\bf u}, 
\ee
yielding, to linear order,
$c_{\rm p}({\bf r})\simeq\bar{c}_{\rm p}(1-\bnabla\cdot{\bf u})$.

To proceed with the profile calculation, we minimize the total free energy with respect to its variables:
the mobile ion densities $n_\pm({\bf r})$, the local fraction of p- and n-type sites $\phi_\pm({\bf r})$, 
the electrostatic potential $\psi({\bf r})$, and the elastic deformation $\bnabla\cdot{\bf u}$.
First, by taking the variation of $F$ in Eq.~\eqref{f_tot} with respect to $n_\pm$, we
obtain the Boltzmann distribution for the concentrations $n_\pm$ 
\bea
n_{\pm} &=& a^{-3} \E^{\beta (\mu_\pm \mp e \psi)}
=n_{\rm b} \E^{\mp \beta e \psi  }
= n_{\rm b}\E^{\mp \Psi},
\eea
where $\beta=1/\kB T$ is the inverse thermal energy, 
$n_{\rm b}=a^{-3} \E^{\beta \mu_\pm}$ and we define the rescaled electrostatic potential
\be
\Psi \equiv {\beta e\psi}.
\label{Psi}
\ee
Furthermore,  because the salt is chosen to be a symmetric monovalent one, we have $\mu_\pm\,{=}\,\mu$.
Minimizing $F$ with respect to $\phi_{\pm}$ yields the p- and n-type charge fractions, which take the form of the Langmuir-Davies isotherm,
\bea
\phi_{\pm} &=& \frac{1}{1 + \E^{\pm \beta e\psi  - A_\pm}} 
=  \frac{1}{1+\E^{\pm\Psi-A_\pm}},
\eea

Upon further minimization of the  free energy $F$ with respect to $\psi({\bf r})$, a modified Poisson-Boltzmann equation
is obtained,
\bea
- \varepsilon \nabla^2 \psi &=& \sum_{i=\pm} q_i n_i 
\nonumber\\
&-& e \bar{c}_{\rm p} ( N_{-}\phi_{-}  -  N_{+}\phi_{+})  
\left( 1 -  \bnabla\cdot {\bf u}\right)  {\mathcal H}(R-r).
\nonumber\\
\label{ELEQ1}
\eea
Finally, the minimization with respect to $\bnabla \cdot {\bf u}$ yields,
\bea
\G  \bnabla \cdot {\bf u} &=&\bar{c}_{\rm p} {\mathcal H}(R-r) \nonumber\\
&\times&
\left[ f_{\rm CR} - e \psi (N_{-}\phi_{-}  -  N_{+}\phi_{+}) - f_{\rm CR}\big|_{\psi{=}0}\right], \nonumber\\
\label{ELEQ2}
\eea
where $f_{\rm CR}\big|_{\psi=0}$ denotes the CR free energy per polymer chain at a reference state, $\psi=0$. 

We choose this zero potential reference state for which the gel is mechanically undeformed, with $\bnabla\cdot{\bf u}=0$. 
The substitution of $f_{\rm CR}\big|_{\psi=0}$ in Eq.~\eqref{ELEQ2} ensures that the CR contribution vanishes at $\psi=0$, 
consistent with zero deformation in the reference state. Note that, for $r>R$, Eq.~\eqref{ELEQ2} formally gives $\bnabla\cdot{\bf u}=0$. 
However, in the model, the polymer network is restricted to the region $r<R$, and only the electrostatic field $\psi({\bf r})$ is non-zero for $r>R$.

The elastic degrees of freedom related to $\bnabla \cdot {\bf u}$ can be eliminated by substituting Eq.~(\ref{ELEQ2}) into Eq.~(\ref{ELEQ1}). 
This means that our charged gel system can be treated as an effective system with only electrostatic degrees of freedom (without elasticity).
In spherical coordinates, we keep only the radial component of the displacement field $u(r)$, 
and the above equations can be written in a decoupled form as
\bea
&-& \varepsilon \frac1{r^2}\frac{\D}{\D r}\Big(r^2 \frac{\D \psi}{\D r}\Big)  \nonumber\\
& =& \sum_{i=\pm} 
q_i n_i  \ - \ e\bar{c}_{\rm p}  (N_{-}\phi_{-} - N_{+}\phi_{+}) {\mathcal H}(R-r) \nonumber\\
&\times & \left( 1 - \frac {\bar{c}_{\rm p} }{\G} \Big[ f_{\rm CR}  -  e  \psi (N_{-}\phi_{-} - N_{+}\phi_{+})-
f_{\rm CR}\big|_{\psi{=}0}  \Big]\right),  \nonumber\\
\label{ELEQ3}
\eea 

\bea
\frac1{r^2}\frac{\D}{\D r}\Big(r^2u\Big)&=& \frac{\bar{c}_{\rm p} }{\G}{\mathcal H}(R-r) \nonumber\\
& \times& \Big[  f_{\rm CR} - 
e\psi (N_{-}\phi_{-} - N_{+}\phi_{+}) - f_{\rm CR}\big|_{\psi{=}0}  \Big]. \nonumber\\
\eea 

The solution of these differential equations yields the equations for the electrostatic potential profile, $\psi(r)$, and 
the local displacement field $u(r)$ 
inside the gel once the boundary conditions are specified. The displacement field $u(r{=}R)$
determines the swollen gel radius, $R \to R+u(R)$, 
and can be viewed as a first-order perturbation to the gel swelling. 
Note that we set the gel boundary in the above equations at its original value, $R$.
Therefore, we neglect the additional effect of the boundary displacement on the electrostatic potential and CR equilibria. 
This approximation applies to small deformations, where $|u(R)|/R\ll1$.

The boundary conditions are imposed as follows. In spherical coordinates, $\psi'(0) \,{=}\, 0$ at the origin, 
as well as $u(0\,){=}\,0$, and 
$\psi'$ denotes the first derivative. 
At the gel perimeter, $r\,{=}\,R$, the electrostatic potential and its first derivative 
should be continuous because the inside and outside dielectric constants are taken to be the same. Namely, 
$\psi(r {\to} R^{-})\,{=}\, \psi(r {\to} R^{+})$, and
$ \psi'(r {\to} R^{-}) \,{=}\, \psi'(r {\to} R^{+})$.
Far from the gel surface, $r{\gg} R$, the electrostatic potential approaches its bulk value, which is taken to be zero, $\psi(\infty)\,{=}\,0$.

\begin{figure}[!t]
	{\includegraphics[width=0.4\textwidth,draft=false]{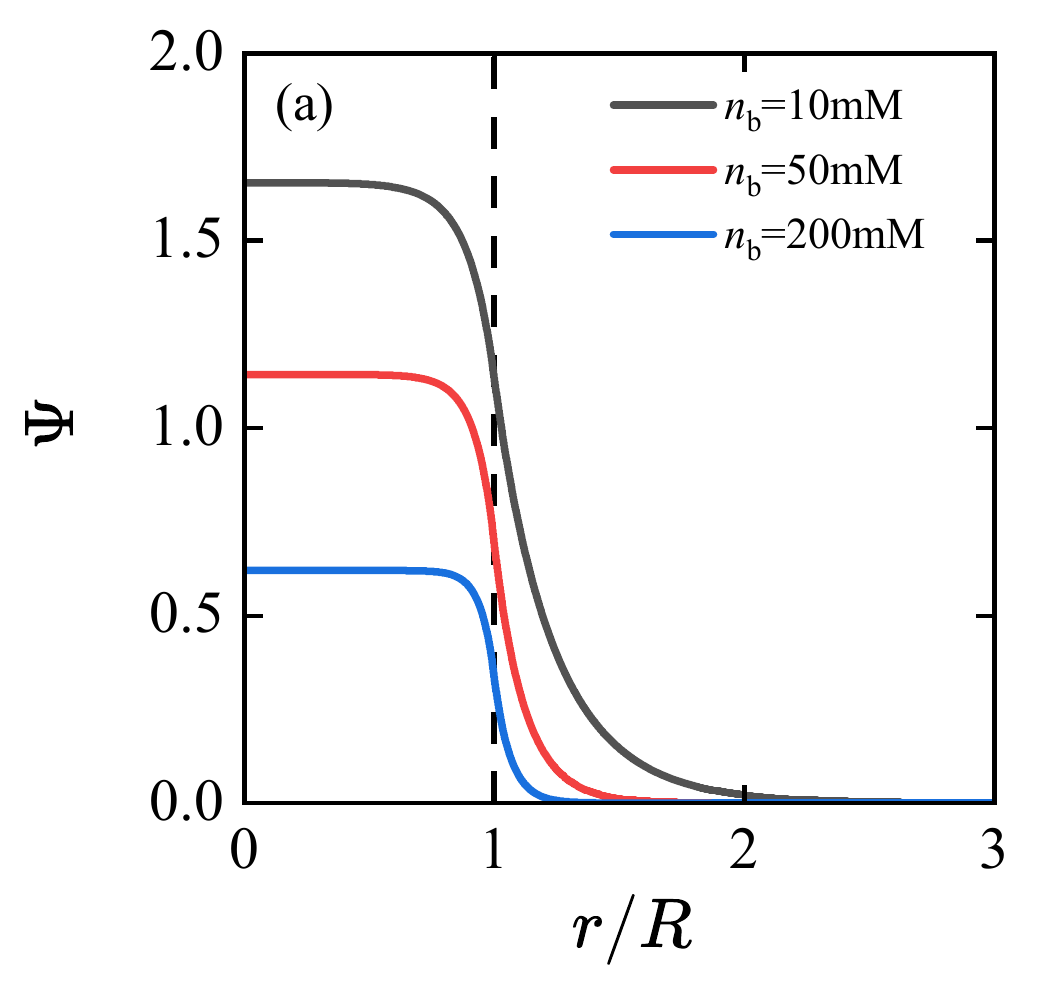}}
	{\includegraphics[width=0.4\textwidth,draft=false]{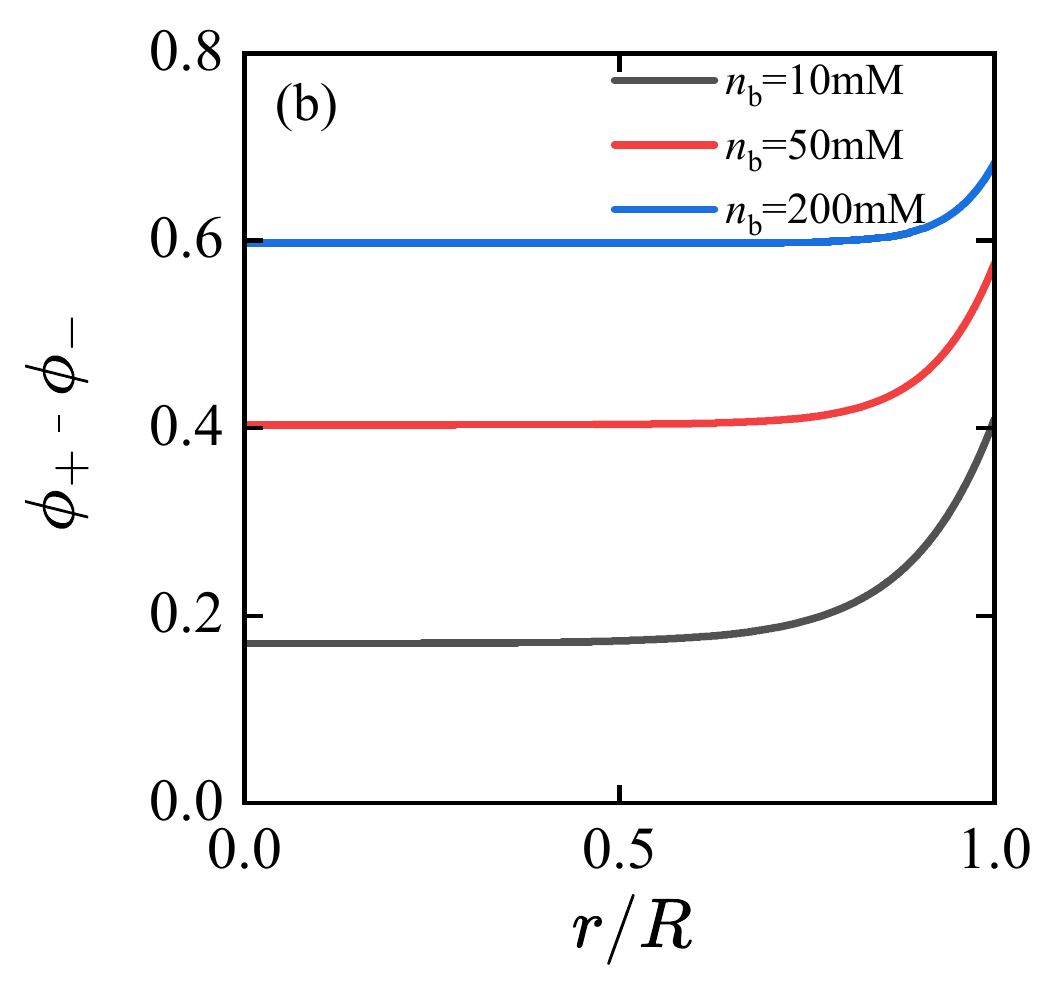}}
	\caption{(a) The rescaled electrostatic potential profile $\Psi{=}\beta e \psi$ as a function of the rescaled distance 
		from the gel center, $r/R$. (b) The spatial distribution of net charge fraction $\phi_{+} - \,\phi_{-}$,
		as a function of $r/R$. The profiles are plotted 
		for different values of salt concentration, $n_{\rm b}\,{=}\,10$\,mM, $50$\,mM, and $200$\,mM. 
		The corresponding Debye lengths are about $3$\,nm, $1.34$\,nm, and $0.67$\,nm, respectively.  
		Thus, the screening length remains comparable to the gel radius $R=10$\,nm, especially at the lowest salt concentration.    
		The other parameters are $A_\pm\,{=}\,\pm2$,
		$N_{\pm}\,{=}\,30$, $R\,{=}\,10$\,nm, and $\bar{c}_{\rm p}/(\beta\G)\,{=}\,0.02$. 
	}
	\label{fig2}
\end{figure}

\section{Results}

We consider a spherical nanogel of radius $R\,{=}\,10$\,nm containing $N_{\rm c}\,{=}\,50$ cross-linked polymer chains. 
Each polymer chain has $N\,{=}\,100$ monomers. These parameters give an estimated polymer volume fraction of about $10\%$ in the gel.
Out of the $N$ monomers, only a fraction $\gamma_{\pm} \,{=}\, 0.3$ 
are p- and n-type, and can undergo association/dissociation, {\it i.e.,} $N_{\pm} \,{=}\, 30$.

We use hereafter the three rescaled variables,
$\Psi$ and $A_\pm$ that were defined earlier, Eqs.~\eqref{Apm} and \eqref{Psi}.
The net gel charge sign can be controlled by adjusting the CR parameters $A_\pm$. 
We focus on a positively charged gel by setting $A_{+} >A_{-}$. 
For comparison, we consider a non-CR gel with a spatially uniform charge density equal to the 
average net charge density of our CR gel at the same salt concentration.
The CR effect will be discussed relative to this non-CR case.

\begin{figure}[!t]
	{\includegraphics[width=0.4\textwidth,draft=false]{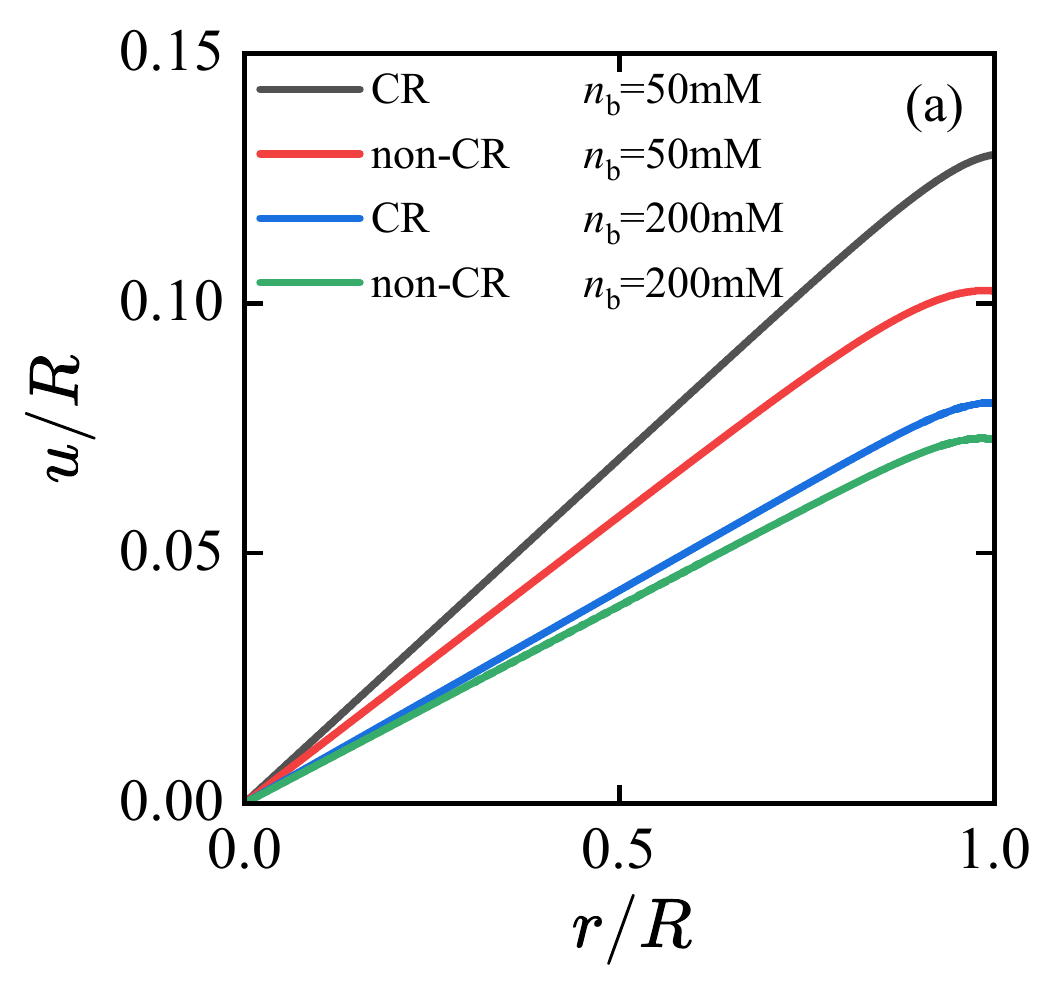}}
	{\includegraphics[width=0.4\textwidth,draft=false]{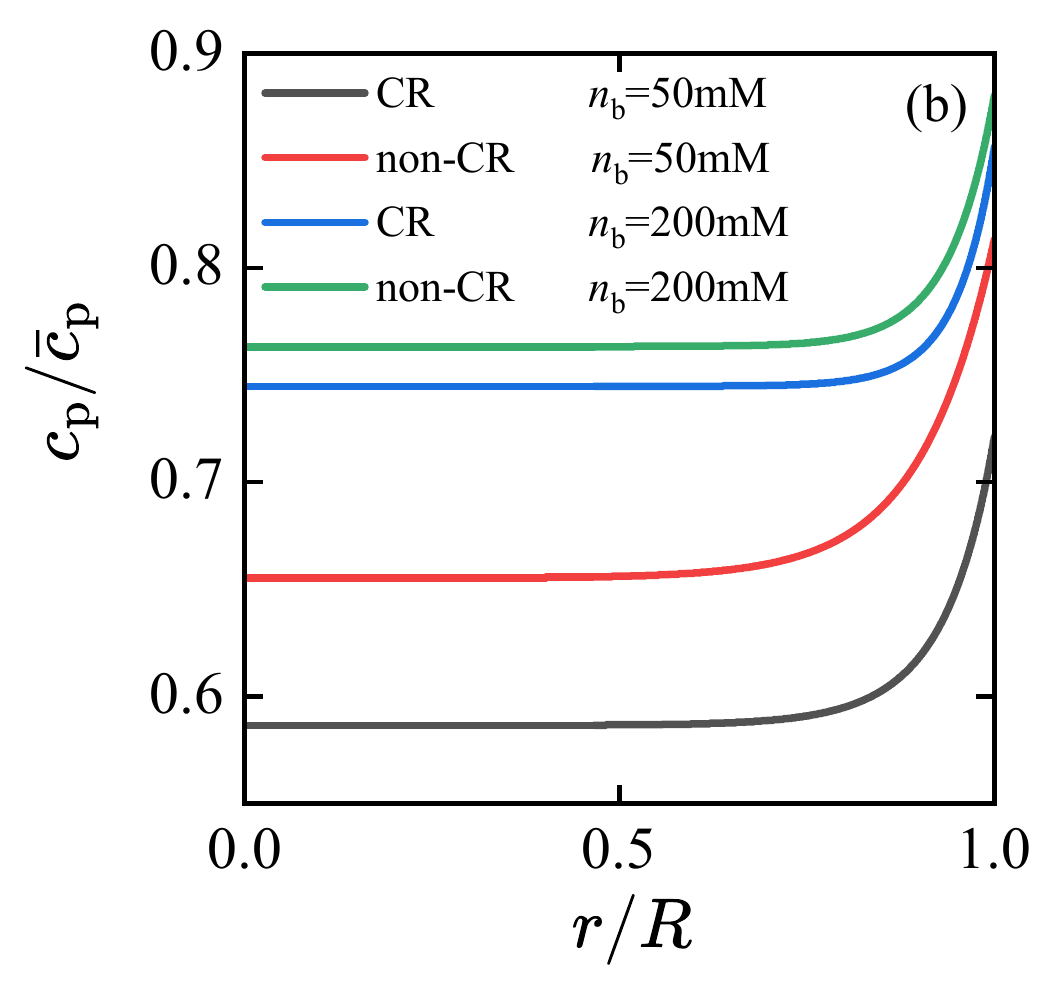}}
	\caption{(a) The spatial distribution of the elastic displacement $u/R$, and (b) the normalized gel density profile, $c_{\rm p}(r)/\bar{c}_{\rm p}$, 
		as a function of the rescaled distance $r/R$ from the gel center. The black ($n_{\rm b}\,{=}\,50$\,mM) 
		and blue ($n_{\rm b}\,{=}\,200$\,mM) lines correspond to the CR case with parameters $A_\pm\,{=} \,\pm 2$, 
		and $N_{\pm}\,{=}\,30$, 
		while the red ($n_{\rm b}\,{=}\,50$\,mM) and green lines ($n_{\rm b}\,{=}\,200$\,mM) correspond to the non-CR case.
		The other parameters are $R\,{=}\,10$\,nm,  and $\bar{c}_{\rm p}/(\beta\G)\,{=}\,0.02$. 
	}
	\label{fig3}
\end{figure}

Figure~\ref{fig2} shows the spatial distribution of the rescaled electrostatic potential $\Psi$ in (a)
and the net charge fraction $\phi_{+} - \phi_{-}$ in the radial $r$ direction in (b). These profiles are plotted for 
three values of salt concentration, $n_{\rm b}\,{=}\,10$\,mM, $50$\,mM, and $200$\,mM.
Both electrostatic potential and net charge fraction exhibit a significant variation 
near the gel interface at $r\,{=}\,R$, while remaining nearly constant in the gel center. 
The addition of salt enhances electrostatic screening, with the corresponding Debye length decreasing from $3.0$\,nm to $1.34$\,nm and $0.67$\,nm for $n_{\rm b}=10$\,mM, $50$\,mM, and $200$\,mM, respectively. 
The stronger screening reduces the electrostatic potential.
Meanwhile, the net charge fraction increases according to the Langmuir--Davies isotherm.

Figure~\ref{fig3}(a) compares the spatial distribution of the elastic displacement $u$ as a function of the rescaled $r/R$,
for the CR  and non-CR cases.
At each salt concentration, the non-CR gel is assigned a uniform charge density  
equal to the volume-averaged net charge density of the corresponding CR gel. As depicted by the black and blue lines, 
$u(r)$ increases monotonically from the gel center ($r\,{=}\,0$) to the gel surface at $r\,{=}\,R$. A spatially dependent $u>0$ 
indicates that the gel is stretched inhomogeneously along the radial $r$-direction. 
The red and green lines show the corresponding profiles for the non-CR cases, and are lower than that for the CR case. 
This suggests that the CR has a strong effect in increasing $u$, thereby enhancing the electro-elastic response. 

Furthermore, increasing $n_{\rm b}$ screens the electrostatic interactions, thereby weakening the electro-elastic response.
Figure~\ref{fig3}(b) presents the normalized gel density profiles $ c_{\rm p}(r)/\bar{c}_{\rm p} $ as a function of $r/R$ for the 
CR and non-CR cases 
at two salt concentrations, $n_{\rm b} \,{=}\, 50$\,mM and $n_{\rm b} \,{=}\,200$\,mM. 
Near the gel center, the concentration remains low and nearly constant. As $r$ increases, $c_{\rm p}(r)/\bar{c}_{\rm p}$ gradually increases 
and reaches its maximum close to the gel surface, indicating a non-uniform swollen state. The lower salt concentration, 
$n_{\rm b} \,{=}\, 50$\,mM yields a smaller overall $c_{\rm p}(r)/\bar{c}_{\rm p}$, and a more pronounced radial variation, 
whereas at $n_{\rm b} \,{=}\, 200$\,mM, 
the profiles are flatter and shifted toward higher density values.

Comparing the CR and the non-CR results, in particular for the lower salt concentration, $n_{\rm b} \,{=}\, 50$\,mM, in Fig.~\ref{fig3}(b)
shows that $c_{\rm p}(r)/\bar{c}_{\rm p}$ in the CR case (black line) 
lies below that in the non-CR case (red line) throughout the gel and varies more strongly with $r$. It indicates that CR enhances 
the non-uniform distribution of the gel density. 

Note that the apparent decrease in the number of monomers within $0\le r\le R$ results from the outward displacement of the gel boundary. 
Upon swelling, the boundary moves to $R+u(R)$, so that part of the polymer network lies in the region $R<r\le R+u(R)$ outside the plotted domain. 
Accordingly, the lower-density curve corresponds to the larger boundary displacement in Fig.~\ref{fig3}(a). 
The total number of monomers remains conserved over the entire deformed gel within the small-deformation approximation.

\begin{figure}[!t]
	{\includegraphics[width=0.4\textwidth,draft=false]{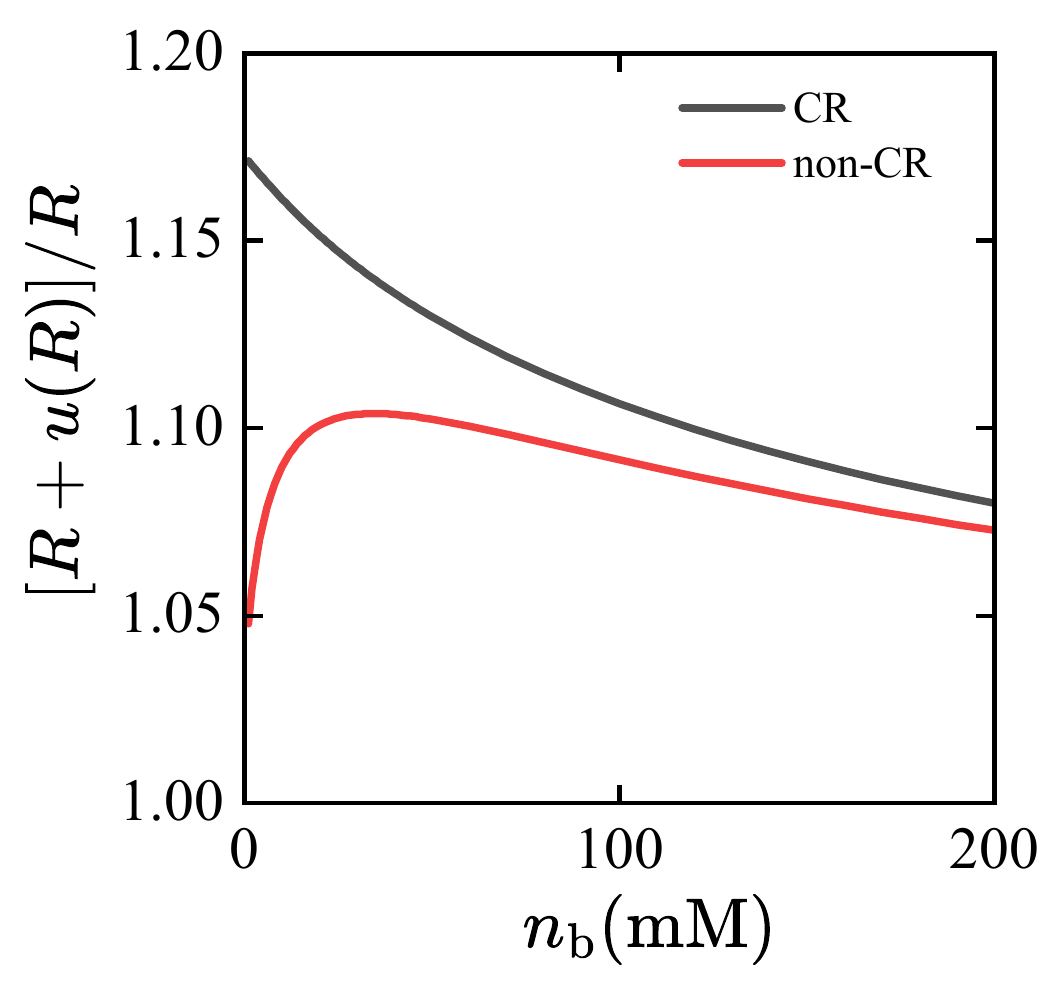}}
	\caption{Salt concentration dependence of the normalized deformed gel radius, $[R+u(R)]/R$, for the CR and non-CR gels. 
		Other parameter values are  $A_\pm\,{=}\,\pm 2$, $N_{\pm}\,{=}\,30$, $R\,{=}\,10$\,nm, 
		and $\bar{c}_{\rm p}/(\beta\G)\,{=}\,0.02$.
	}
	\label{fig4}
\end{figure}

The effect of CR on the overall gel size is further shown in Fig.~\ref{fig4}, where the normalized deformed gel radius, $[R+u(R)]/R$, 
is plotted as a function of the salt concentration. For the CR case, the gel size decreases with increasing $n_{\rm b}$ because stronger 
electrostatic screening weakens the electro-elastic deformation.  In contrast, the non-CR gel size first increases 
with $n_{\rm b}$ at low salt concentrations and then decreases at higher $n_{\rm b}$, showing a non-monotonic dependence on $n_{\rm b}$.
Moreover, the CR gel is larger than the corresponding non-CR gel, particularly at low salt concentrations. 
Note that the fixed charge density of the non-CR gel varies with $n_{\rm b}$ to match the corresponding CR charge density.

\section{Discussion and Summary}

We have studied the effects of CR on the non-uniform swelling behavior of polyelectrolyte hydrogels. 
Our model combines the Poisson-Boltzmann theory, a two-site CR model, 
with an electro-elastic coupling. 
A set of coupled differential equations governing the electrostatic potential and the elastic displacement is derived, 
and reveals a balance between the elastic and electrostatic forces. 

Our study uses two approximations. (i) The polymer density is linearly coupled to the elastic deformation 
through $c_{\rm p}({\bf r})=\bar{c}_{\rm p}(1- \bnabla\cdot{\bf u})$. Nonlinear density-deformation effects are not considered.
(ii) The coupled equations are solved by keeping the gel boundary at its undeformed value $R$, while the deformed radius 
is estimated as $R+u(R)$. Hence, the added effect of the displaced boundary on the electrostatic potential and CR equilibria is neglected.
Both approximations are based on the small-deformation assumption. 
The present formulation can be extended by treating the displaced boundary self-consistently and by 
retaining the nonlinear relation between the polymer density and deformation. 

We show that CR significantly increases non-uniform swelling by generating 
stronger spatial variations in the charge distribution within the gel, as compared with the non-CR case.
Increasing the salt concentration weakens electrostatic interactions, lowers the electrostatic potential.
Although the net charge increases, the stronger electrostatic screening dominates and reduces the elastic displacement.

The non-uniform swelling in the present work originates from the electro-elastic coupling mechanism. 
This can be seen by rewriting Eq.~(\ref{ELEQ2}) as a local force-balance relation,
\be
\G \bnabla(\bnabla \cdot {\bf u} )  = -e \bar{c}_{\rm p} {\mathcal H}(R-r)  
( N_{+}\phi_{+} - N_{-}\phi_{-}) {\bf E} ,
\label{force_balance}
\ee 
where the electric field is ${\bf E}=-\bnabla\psi$. 
The above equation shows that the spatial variation of $\bnabla\cdot{\bf u}$ is induced by the electrostatic 
body force acting on the charged polymer network, 
as it depends on the coupling term, $(N_{+}\phi_{+} \,{-}\, N_{-}\phi_{-}){\bf E}$. 
Thus, the deformation is governed not only by the total gel charge, but also by the radial distribution of the charge fraction 
and electric field. Although the gel remains spherically symmetric, these quantities vary with $r$, 
resulting in a nonuniform swelling response along the radial $r$ direction.

The CR mechanism enhances this trend by making $\phi_{\pm}$ responsive to the local electrostatic potential as shown in Fig.~\ref{fig2}(b), 
unlike the non-CR case with fixed charge fractions. 
As a result, the variation of $\psi(r)$ produces a spatially varying net charge fraction, $\phi_{+}(r) \,{-}\, \phi_{-}(r)$, 
which leads to a larger displacement than that of the matched non-CR case.

The polymer density profile in Fig.~\ref{fig3}(b) follows from the elastic displacement field through 
$c_{\rm p}({\bf r})\,{=}\,\bar{c}_{\rm p}(1- \bnabla{\cdot}{\bf u})$.
Near the gel center, $\bnabla{\cdot}{\bf u}$ remains large and nearly constant, yielding a low polymer 
density and hence a more swollen inner region.
Toward the gel boundary, $\bnabla{\cdot}{\bf u}$ varies more strongly due to the stronger radial variations 
of the electric field and charge fraction. 

The different salt dependence of the CR and non-CR gel sizes 
in Fig.~\ref{fig4} can be understood as arising from the competition between charge regulation and electrostatic screening. 
For the non-CR gel, increasing $n_{\rm b}$ at low salt concentration increases the uniform charge assigned from 
the corresponding CR gel and enhances $u(R)$. At higher $n_{\rm b}$, 
electrostatic screening becomes stronger and suppresses deformation, 
leading to a non-monotonic variation in the gel size. 

The CR gel shows a different behavior because its deformation is governed by the full CR-electrostatic contribution 
in Eq.~\eqref{ELEQ2}, rather than by the net charge alone. 
As $n_{\rm b}$ increases, both $\Psi$ and its spatial variation decrease. The resulting weaker electro-elastic response 
leads to a monotonic decrease of the gel size.

One of the findings is that the polymer density increases outward, giving a non-uniform profile that evolves from a more swollen 
inner region to a relatively denser outer region. Compared with the non-CR case, CR further strengthens 
this density heterogeneity by producing lower $c_{\rm p}(r)$ 
and a steeper radial profile, especially at low salt concentration. Increasing the salt concentration 
suppresses the CR effect through electrostatic screening, 
thereby reducing the density contrast and approaching a more uniform profile.

Our theoretical model focuses on nanogels for which the gel radius is comparable to the Debye screening length. 
In our calculations, the Debye screening length ranges from about $3$\,nm to $0.67$\,nm, with its largest value reaching nearly one third of the gel radius.
In this regime, the spatial coupling among electrostatics, charge regulation, and elastic deformation becomes important.

In conclusion, our study provides a theoretical framework for 
understanding how charge regulation enhances nonuniform swelling in nanogels through electro-elastic coupling. 
The results may provide useful guidance for the tailored design and optimization of functional hydrogel devices.

\begin{acknowledgments}
This article is dedicated to the memory of our dear friend and colleague, Rudolf (Rudi) Podgornik, 
who contributed to the early stages of this work and passed away unexpectedly in December 2024. 
A pioneering scientist in soft matter physics, 
Rudi left an enduring mark on the field through his profound scientific contributions and remarkable intellectual generosity. 
He will be greatly missed and fondly remembered.
	
B.Z. thanks Fangfu Ye for the support on computing resources, and acknowledges the National Natural Science Foundation of China (NSFC) through Grants No. 22203022 and the Scientific Research Starting Foundation of Wenzhou Institute, UCAS (No.~WIUCASQD2022016). 
L.H. acknowledges the support from the National Natural Science Foundation of China (Grant No.~22273067).
X.W. acknowledges the support from the National Natural Science Foundation of China (Grant No.~12374215).
S.K.\ acknowledges the support by the National Natural Science Foundation of China (Grant No.\ 12274098) and 
by the Zhejiang Key Laboratory of Soft Matter Biomedical Materials (2025ZY01036 and 2025E10072).
D.A. acknowledges partial support from the Israel Science Foundation (ISF) under grant No. 226/24.
\end{acknowledgments}
\newpage

\end{document}